\documentclass[aps,11pt]{revtex4}
\usepackage{graphicx}
\usepackage{bm}
\begin{document}
\title{El Ni$\tilde{\mbox{n}}$o signature in Alaskan river breakups}
\author{G.~Boffetta}
\affiliation{
Dipartimento di Fisica Generale and INFN,
Universit\`a degli Studi di Torino, v. Pietro Giuria 1, 10125, Torino, Italy \\
and CNR-ISAC, Sezione di Torino, c. Fiume 4, 10133 Torino, Italy}

\begin{abstract}
A signature of El Ni$\tilde{\mbox{n}}$o-Southern Oscillation
is found in the historical dataset of the Alaskan Tanana river breakups
where the average ice breaking day is found to anticipate of about 
$3.4$ days when conditioned over El Ni$\tilde{\mbox{n}}$o years. 
This results represents a statistically significant example of ENSO 
teleconnections on regions remote from tropical Pacific.
\end{abstract}

\maketitle

The El Ni$\tilde{\mbox{n}}$o-Southern Oscillation (ENSO) is the most
important recurrent pattern in the interannual climate variability.
It corresponds to anomalous warm water in the tropical Pacific
ocean (El Ni$\tilde{\mbox{n}}$o) coupled to an atmospheric pattern 
over Indian and Pacific oceans which strongly reduces trade winds.
Basic physical ingredients for the appearance of El Ni$\tilde{\mbox{n}}$o,
occurring irregularly every $2$ to $7$ years, are now understood,
but 
the extension of its impact is still partially unknown \cite{MZG06}.
The effects of El Ni$\tilde{\mbox{n}}$o are more
consistent in the tropical Pacific areas.
At higher latitudes and in remote areas the impact is less
deterministic: a single event is mediated by local meteorological
conditions and ENSO-related variations are better characterized in 
a statistical sense.
Recent statistical studies have reported effects on a wide spectrum 
of human and natural activities via the so-called teleconnections. 
To cite some examples, ENSO signature has been found in forest fire
statistics, in disease epidemics and in financial market \cite{MZG06}.
Here we report the analysis of ENSO teleconnections on the 
date of ice breakup in Tanana river in Nenana, Alaska.
Thanks to the rather long time dataset ($90$ years) it is possible to
disentangle the imprint of ENSO signal on the local meteorological noise
with good statistical significance.

The Tanana data set is an non-conventional time series obtained from
the Nenana Ice Classis lottery. Since 1917, participants to the lottery have to
guess the day (and minute) of ``official'' Tanana river breaking,
defined as the time at which a wooden tripod frozen into the river
moves and stops a clock mechanism at the shore \cite{NIC}. 
Nenana is still a small town of about 500 people, and the tripod has 
been placed in about the same section of the
river, therefore we can assume that local conditions are not much changed
in the $90$ years of record. 
Moreover, ice breaking in rivers is a complex phenomenon due to both 
direct thermal (local
temperature) and mechanical (snow melting upstream) effects
and therefore it is a good indicator of the seasonal climate of the
region. For this reason, ice breaking in rivers and lakes has been used 
to infer trends in 
climate variation over the past centuries \cite{Metal00}. 
In particular, the analysis of Tanana dataset revealed a negative 
(anticipating) trend in the breakup day of about $5.5$ days \cite{SM01}.
Our observation of ENSO modulations in the Tanana dataset gives a 
support to the climatological origin of the observed trend.
The $90$-year record of Tanana (Fig.~1A) shows large variability with
a breakup day ranging from April 20 (in 1940 and 1998) (day $d=110$ 
of the year) to May 20 (in 1964) ($d=140$).
Without taking into account trends (but correcting for leap years), 
the mean breakup day is $<d>=124.7$, corresponding to May 5th. The 
underlying distribution is rather wide, with a variance 
$\sigma_{d}=5.97$ days.

El Ni$\tilde{\mbox{n}}$o years are classified by a number of different
criteria. Mostly used are the an oceanic index based on Sea Surface Temperature
(SST) anomalies, and the atmospheric Southern Oscillation Index (SOI)
based on pressure difference between Tahiti and Darwin \cite{MZG06}.
The present analysis makes use of the NOAA Bivariate EnSo Timeseries (BEST) 
index \cite{BEST} which is a combination of Ni$\tilde{\mbox{n}}$o 3.4 SST and
SOI with a 5-month running mean. El Ni$\tilde{\mbox{n}}$o years
are defined as those for which the Best index is positive for at 
least four months between previous October and April.
Changes in the details of this definition do not alter significantly the 
results. According to this definition, there have been $14$ El 
Ni$\tilde{\mbox{n}}$o years between $1917$ and $2006$, the most recent
ones in $2002-03$ and in $1997-98$. In particular, the $1997-98$ was 
a strong El Ni$\tilde{\mbox{n}}$o winter followed by one of the two
earliest breakup in Tanana (the other, in 1940, also followed an 
El Ni$\tilde{\mbox{n}}$o winter). This coincidence suggests possible
connections between ENSO and ice breaking in Tanana river.

We have analyzed the correlation between Tanana dataset and Best index
by computing the mean breaking day conditioned to El Ni$\tilde{\mbox{n}}$o
years. The resulting $<d^*>=121.3$ significantly differs from the 
unconditioned mean by a $3.4$ day difference.
In order to give a statistical meaning to the observed anticipation in
breakup, we can make the null hypothesis of no correlation and 
perform a simple test based on 
Monte Carlo surrogate data. By taking $N=14$
random years uniformly distributed between $1917$ and $2006$, we compute
the mean breaking day $\bar{d}$ conditioned over these years. By definition
the average value over many realizations is again $<\bar{d}>=124.7$ with
a distribution very close to a Gaussian (thanks to the Central Limit
Theorem) with variance of the mean 
$\sigma_{\bar{d}} = \sigma_{d}/\sqrt{N} \simeq 1.6$.
Conditioning over El Ni$\tilde{\mbox{n}}$o therefore
gives a mean breakup day $<d^*>$ 
which is at $2.3$ standard deviations below the mean.
The probability that such a small value has been obtained by chance
is only of $1.1 \%$ and we can refute the null hypothesis (no correlation) 
with a $98.9 \%$ of confidence (see Fig.~1B).

Both temperature and snowfall contribute to the breaking of ice 
in a complex way. In the case of lakes, the response
to temperature variations is known to be strongly non-linear \cite{WML04},
while the presence of snow has less known net effect as from one side 
protects ice from melting but, from the other side, increases the 
water flow below.
In order to get more insight on the physical mechanisms of the 
breakup, we have analyzed the meteorological data of the Alaska
Climate Research Center of the University of Alaska
at Fairbanks \cite{ACRC}. 
This remarkable dataset gives the minimum and maximum air
temperature, precipitation, snowfall and snow depth with daily
resolution for $77$ years (1930-2006) in the town of Fairbanks, 
about $90$ km upstream Nenana.
We have found that both maximal temperature and snowfall (and snowdepth)
are statistically affected by ENSO. The average maximal temperature
of the first $120$ days (Jan-Apr) rises from $-7.6~^{o}C$ to
$-6.3~^{o}C$ if conditioned over the $12$ El Ni$\tilde{\mbox{n}}$o
events of the period while the total snowfall on the same period
decreases during El Ni$\tilde{\mbox{n}}$o years from an average 
$75~cm$ to $55~cm$.
The probabilities that these variations are due to random fluctuations
are comparable and, when evaluated with MonteCarlo surrogates,
are around $3 \%$. A direct consequence of reduced
snowfall and increased temperature during ENSOs is the acceleration of 
snow melting on ground, which is one of the known mechanisms for ice breaking.
Indeed, we have found that ice breaking days in Nenana are strongly
correlated with the "no-snow" days in Fairbanks (defined as the first day 
without snow on ground). Indeed the correlation coefficient between
the two dates on the whole $90$ years dataset is $\rho\simeq 0.50$.

In conclusion, the breakup of Alaskan Tanana river is shown to be 
influenced by El Ni$\tilde{\mbox{n}}$o condition in the previous winter.
This result supports the extension of ENSO teleconnections to regions 
remote from tropical Pacific. At these latitudes effects are 
significant only in a statistical sense and therefore can be resolved 
on the bases of sufficiently long dataset. 
Unfortunately, their
knowledge is of little use for betting in the next Nenana lottery.

\begin{figure}[h!]
\includegraphics[draft=false,scale=0.6]{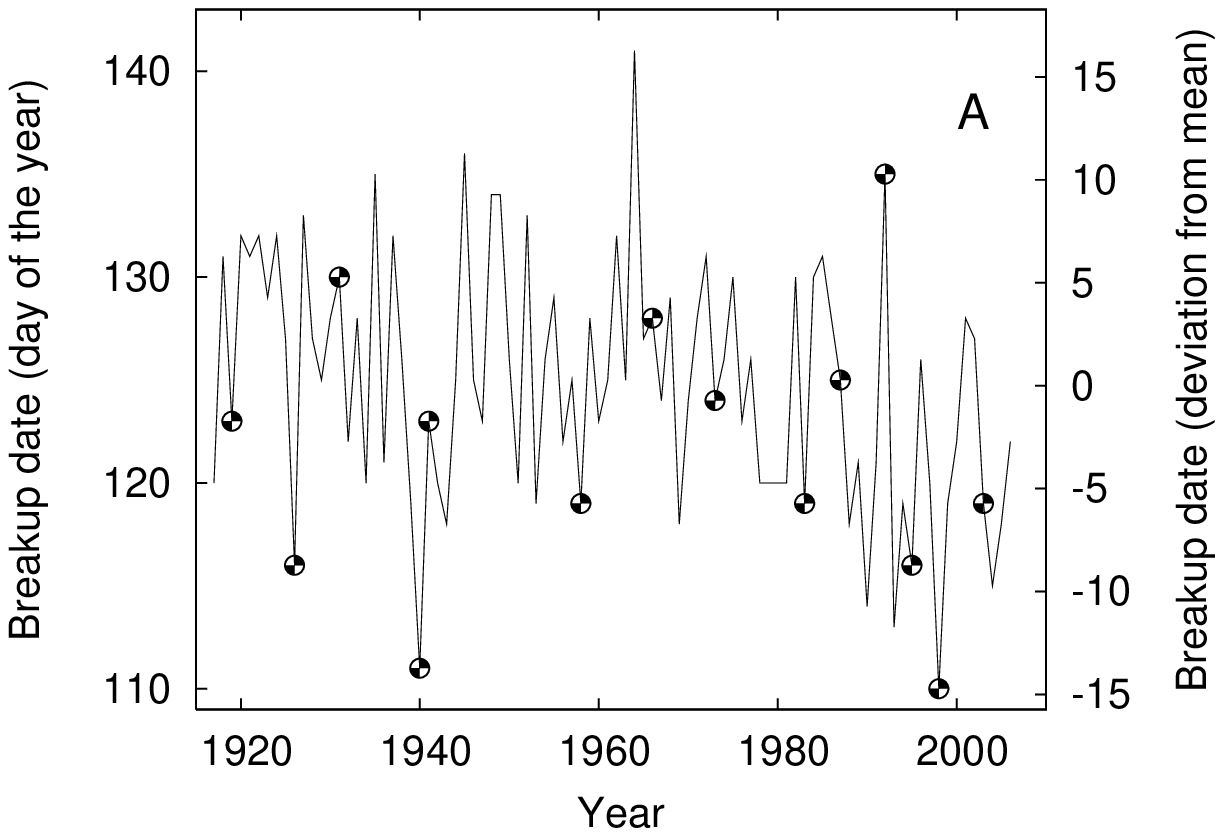}
\includegraphics[draft=false,scale=0.6]{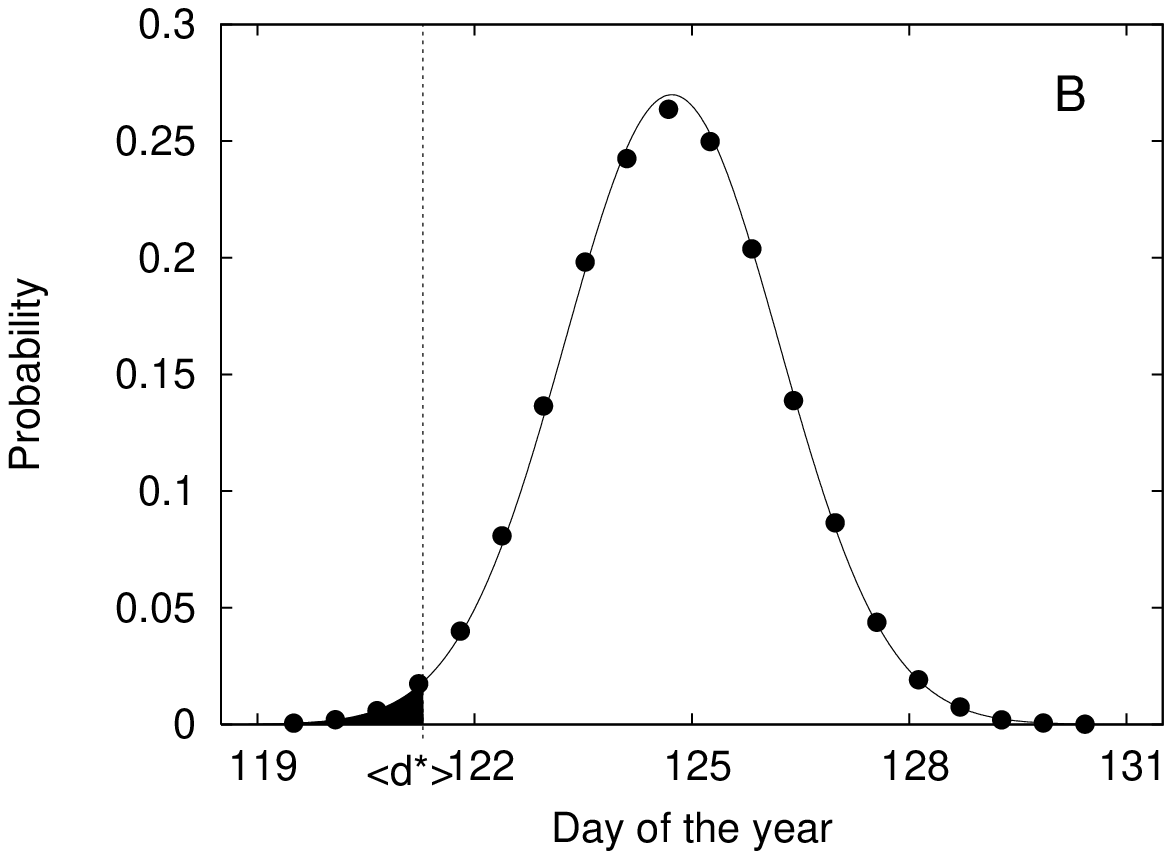}
\caption{(A) Breakup date of the Tanana river from the Nenana Ice
Classic lottery. Points correspond to El Ni$\tilde{\mbox{n}}$o years.
(B) Probability density function of $10^{6}$ MonteCarlo realizations
of the mean breaking day $\bar{d}$ averaged over $N=14$ random 
years in the range
1917-2006. Line represents a Gaussian with the same mean and
variance. The vertical line indicates the mean breaking day conditioned over
ENSO years, $<d^{*}>=121.3$. The probability of a random realization of these
years is given by the black area left to the line and is equal to
$1.1 \%$.  }
\label{fig1}
\end{figure}


\end{document}